\documentclass[useAMS,usenatbib]{mn2e}

\title[Quasar Clustering in Cosmological Hydrodynamic
  Simulations]{Quasar Clustering in Cosmological
  Hydrodynamic Simulations: Evidence for mergers}

\author[Colin Degraf et al.]  {Colin Degraf$^{1}$, Tiziana Di Matteo$^{1}$, Volker Springel$^{2}$\\ $^{1}$ McWilliams Center for
       Cosmology, Carnegie Mellon University, 5000 Forbes Avenue, Pittsburgh,
       PA 15213, USA\\ $^{2}$ Max-Planck-Institut f\"{u}r
       Astrophysik, Karl-Schwarzschild-Stra\ss e 1, 85740 Garching bei
       M\"{u}nchen, Germany\\}

\usepackage{ graphicx}
\usepackage{ subfigure}
\usepackage{multirow}

\begin{document}

\date{Accepted 200? ???? ??.
      Received 2008 ???? ??;
      in original form 2008  xx}
\pagerange{\pageref{firstpage}--\pageref{lastpage}}
\pubyear{200?}

\maketitle

\begin{abstract}
  We examine the clustering properties of a population of quasars drawn from
  fully hydrodynamic cosmological simulations that directly follow black hole
  growth.  We find that the black hole correlation function is best described
  by two distinct components: contributions from BH pairs occupying the same
  dark matter halo ('1-halo term', $\xi_{\rm{BH,1h}}$) which dominate at
  scales below $\sim 300 \: \rm{kpc \: h^{-1}}$, and contributions from BHs
  occupying separate halos ('2-halo term', $\xi_{\rm{BH,2h}}$ ) which dominate
  at larger scales.  From the 2-halo BH term we find a typical host halo mass
  for faint-end quasars (those probed in our simulation volumes) ranging from
  $M\sim 10^{11}$ to a few $10^{12} M_{\odot}$ from $z=5$ to $z=1$
  respectively (consistent with the mean halo host mass). The BH correlation
  function shows a luminosity dependence as a function of redshift, though
  weak enough to be consistent with observational constraints.  At small
  scales, the high resolution of our simulations allows us to probe the 1-halo
  clustering in detail, finding that $\xi_{\rm{BH,1h}}$ follows an approximate
  power law, lacking the characteristic decrease in slope at small scales
  found in 1-halo terms for galaxies and dark matter.  We show that this
  difference is a direct result of a boost in the small-scale quasar bias
  caused by galaxies hosting multiple quasars (1-subhalo term) following a
  merger event, typically between a large central subgroup and a smaller,
  satellite subgroup hosting a relatively small black hole. We show that our
  predicted small-scale excess caused by such mergers is in good agreement
  with both the slope and amplitude indicated by recent small-scale
  measurements.  Finally, we note the excess to be a strong function of halo
  mass, such that the observed excess is well matched by the multiple black
  holes of intermediate mass ($10^7-10^8$ M$_{\odot}$) found in hosts of $M
  \sim 4-8 \times 10^{11} M_\odot$, a range well probed by our simulations.

\end{abstract}

\section{Introduction}
With supermassive black holes being found at the centre of most galaxies
\citep{1995ARA&A..33..581K}, interest in quasars has increased significantly,
with substantial investigation into fundamental relations between black hole
masses and their host galaxies' properties \citep{1998AJ....115.2285M,
  2000ApJ...539L...9F, 2000ApJ...539L..13G, 2002ApJ...574..740T,
  2007ApJ...655...77G}.  In addition to these relations, statistical studies
of the spatial clustering of quasars provide the potential to better
understand the relation between quasars, their hosts and the underlying dark
matter distribution, as well as estimate quasar lifetimes \citep[see,
e.g.,][]{HaimanHui2001, MartiniWeinberg2001} across a relatively large range
of redshift. For example, strong clustering would suggest quasars should
reside in massive groups. If so, they should be rare and in order to reproduce
the quasar luminosity density, they must have long lifetimes. Conversely,
low correlation would suggest more common quasars, and thus shorter quasar
lifetimes.

Early studies of quasar clustering produced varying results for the clustering
amplitude, with no clear agreement on overall evolution with redshift, some
suggesting minimal or decreasing clustering evolution \citep{MoFang1993,
  CroomShanks1996}, while others found an increase in clustering with redshift
\citep{Kundic1997, LaFranca1998}. These findings were generally poorly
constrained due to the small sizes of available quasar samples. With the
emergence of large scale surveys such as Sloan Digital Sky Survey
\citep{2000AJ....120.1579Y} and the Two-degree Field QSO Redshift Survey
\citep{2002MNRAS.333..279L}, substantially larger catalogs have been compiled,
permitting more detailed investigation into the clustering properties of
quasars, and many recent studies have been made into this area
\citep[e.g.][]{LaFranca1998, Porciani2004, Croom2005, Shen2007, Myers2007,
  daAngela2008, Shen2009, Ross2009}. These recent studies have found evidence
for an increase in clustering amplitude with redshift \citep{LaFranca1998,
  Porciani2004}, primarily for $z>2$, in agreement with predictions from
simulations \citep[see, e.g.][]{Bonoli2009, Croton2009}.

In addition to overall evolution, the luminosity dependence (if any) of
large-scale clustering can provide significant insight into what quasar
populations dominate different luminosity ranges. For example, the model of
\citet{2005ApJ...630..705H, 2005ApJ...630..716H, 2005ApJ...632...81H,
  2005ApJ...625L..71H, 2006ApJS..163....1H} suggests that bright and faint
quasars are similar objects which are observed at different phases of their
lifetimes, rather than being fundamentally different populations of quasars
(as simpler, 'on-off' models assume). This model would suggest that both
bright and faint quasars should populate similar halos. Thus, while there may
be some correlation between peak luminosity and host halo mass, clustering
dependence on instantaneous luminosity should be relatively weak, particularly
when compared to more traditional 'on-off' models of quasar luminosity
\citep{Lidz2006}. Recent observational studies have generally found a lack of
luminosity dependence in the correlation function \citep[see,
e.g.,][]{Croom2005, Myers2007, daAngela2008}, though \citet{Shen2009} found
evidence for some, though weak, luminosity dependence.  Several semi-analytic
models have also been used, finding differing luminosity dependences, such as
a significant dependence for sufficient luminosity ranges, but limited when
considering only luminosities probed by observation \citep{Bonoli2009}, or
weak dependence at low redshift ($z<1$), but stronger at higher redshift
\citep{Croton2009}. 

In addition to large scale behavior, the possibility of excess quasar
clustering on very small scales has arisen in several recent studies.  While
some observed quasar pairs are believed to be the result of gravitationally
lensed quasars, it has been proposed that others may be physically distinct
quasar binaries, which would suggest quasars cluster much more strongly on
small scales than extrapolation of large scale clustering would imply
\citep{Djorgovski1991, Hewett1998, Kochanek1999, Mortlock1999}, suggesting a
connection between galaxy mergers and quasar activity \citep[see,
e.g.][]{Kochanek1999}.  However, investigating the smallest scale clustering
has typically been problematic due to observational limitations (such as fiber
collisions preventing small-separation pairs from being distinguished as
distinct objects) and sample sizes insufficient for probing the smallest
scales, where quasar pairs are rare.  There have been several studies probing
clustering at sub-Mpc scales, generally finding no excess clustering relative
to an extrapolation of the large-scale clustering behavior \citep[see,
e.g.][]{Shen2009b, Padmanabhan2009}.  However, these studies have been limited
to scales above 100 $\rm{kpc \: h^{-1}}$, while several recent studies have
managed to probe even smaller scales, where they do indeed find a significant
excess \citep{Hennawi2006, Myers2007II, Myers2008}.

In particular, \citet{Hennawi2006} studied binary quasars from SDSS and 2dF
Quasar Survey to compute quasar clustering for scales as small as 20 $\rm{kpc}
\: h^{-1}$ (comoving), and found significant excess clustering relative to the
large scale extrapolation (by an order of magnitude at comoving scales below
100 $\rm{kpc} \: h^{-1}$, and growing stronger with decreasing scale).  This
excess implies that the quasars are more strongly clustered than galaxies at
these small scales, supporting the theory that quasar activity is triggered by
galaxy interactions.  Using the quasar sample from \citet{Myers2007},
\citet{Myers2007II} found only a slight excess in small-scale clustering, and
put an upper limit for the excess at a factor of 4.3\underline{+}1.3 for
physical scales of $\sim 28 \: \rm{kpc} \: h^{-1}$.  They suggest that the
significantly larger excess of \citet{Hennawi2006} is a result of a selection
effect, possibly due to studies tending to target tracers of the Ly$\alpha$
forest, causing a bias toward $z > 2$, which may be more highly clustered.
\citet{Myers2008} used a complete spectroscopic sample of quasars over
physical scales of 23.7-29.9 $\rm{kpc} \: h^{-1}$ from SDSS to find an excess
clustering factor of $\sim$4, consistent with the upper limit of
\citet{Myers2007II}, which, while 2$\sigma$ below the excess found by
\citet{Hennawi2006}, nonetheless supports the general finding of a clustering
excess which may be a result of galaxy interactions.

In this paper, we use cosmological hydrodynamic simulations which directly
model the growth, accretion, and feedback processes of black holes to
investigate the properties and underlying causes of black hole clustering.
Although the simulation volume limits our analysis to black hole luminosities
and host group masses below those typically studied, the self-consistent
modeling of black holes allows us to study the clustering behavior without
post-processing models.  Additionally, the high resolution allows us to
investigate clustering behavior at extremely small scales, well below those
studied with semi-analytic models, thereby providing a means of using
simulations to investigate the observed small-scale excess for the first time,
and provide a physical explanation for the underlying cause.

In Section 2 we describe the numerical modeling for the black holes formation
and accretion (Section 2.1) the simulation parameters used (Section 2.2), the details of the subgroup finder (Section 2.3) and our method of calculating
correlation functions (Section 2.4).  In Section 3 we investigate the quasar
clustering properties at both large and small scales, and we summarize our
results in Section 4.

\begin{figure*}
\centering
\includegraphics[width=14cm]{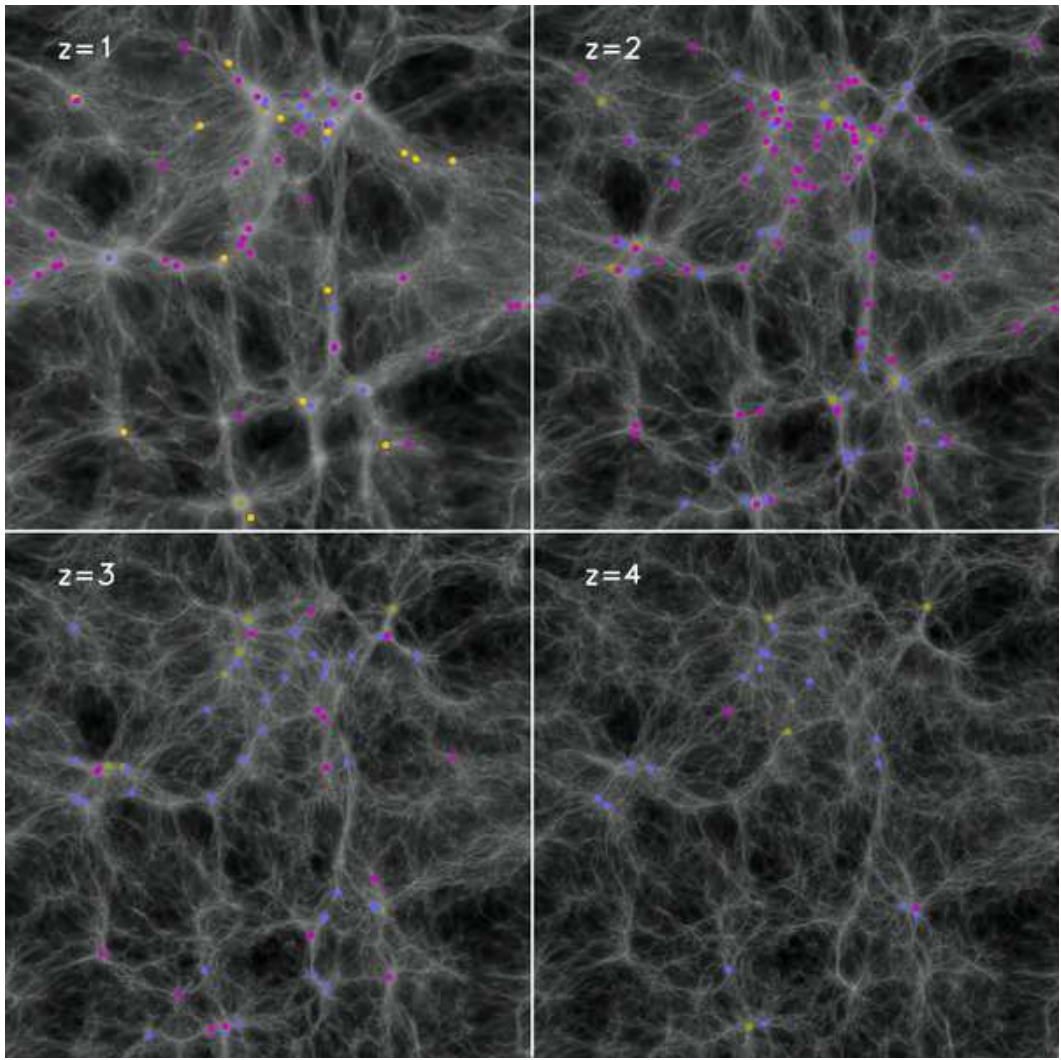}
\caption{An example of the distribution of black holes in the simulations: The
  same slice ($2 \: \rm{Mpc} \: h^{-1}$ thick) through the D6 simulation at
  z=1,2,3,4. The positions of black holes in different luminosities bins ($L <
  10^8 L_\odot$ - Orange; $10^8 L_\odot < L < 10^9 L_\odot$ - Pink; $10^9
  L_\odot < L < 10^{10} L_\odot$ - Blue; $L > 10^{10} L_\odot$ - Green.) are
  plotted on top of the gas density distribution (shown in the the gray
  scale).}
\label{simulationslice}
\end{figure*}

\section{Method}

\subsection{Numerical simulation}

In this study, we analyse the set of simulations published in
\citet{DiMatteo2008}. Here we present a brief summary of the simulation
code and the method used. We refer the reader to \citet{DiMatteo2008}
for all details.

 The code we use is the massively parallel cosmological TreePM--SPH code
 {\small Gadget2} (Springel 2005), with the addition of a multi--phase
 modeling of the ISM, which allows treatment of star formation (Springel \&
 Hernquist 2003), and black hole accretion and associated feedback processes
 (Springel et al. 2005, Di Matteo et al. 2005).

Black holes are simulated with collisionless particles that are created in
newly emerging and resolved groups/galaxies.  To find these groups, a
friends--of--friends group finder is called at regular intervals on the fly
(the time intervals are equally spaced in log $a$, with $\Delta \log{a} =
\log{1.25}$), finding groups based on particle separations below a specified
cutoff.  Each of these groups that does not already contain a black hole is
provided with one by turning its densest particle into a sink particle with a
seed black hole of fixed mass, $ M = 5 \times 10^5 h^{-1}$\,M$_\odot$. After insertion,
the black hole particle grows in mass via accretion of surrounding gas
according to $\dot{M}_{\rm BH} = \frac {4 \pi G^2 M_{\rm BH}^2 \rho}{(c_s^2 +
  v^2)^{3/2}}$ \citep{1939PCPS...35..405H, 1944MNRAS.104..273B,
  1952MNRAS.112..195B}, and by merging with other black holes.  Note that
within the simulations, it is assumed that accretion is limited to a maximum
of 3 times the Eddington rate, although very few sources accrete above
$\dot{M}_{Edd}$.

	The accretion rate of each black hole is used to compute the
	bolometric luminosity, $L = \eta \dot{M}_{\rm BH} c^2$
	\citep{1973A&A....24..337S}.  Here $\eta$ is the radiative efficiency,
	and it is fixed at 0.1 throughout the simulation and this analysis.
	Some coupling between the liberated luminosity and the surrounding gas
	is expected: in the simulation 5 per cent of the luminosity is
	(isotropically) deposited as
	thermal energy in the local black hole kernel, acting as a form of feedback energy \citep{2005Natur...433..604D}.

\subsection{Simulation parameters}
\begin{table}
\caption{Numerical Parameters}
\begin{tabular}{c c c c c c}

  \hline
  \hline
  
  Run & Boxsize & $N_p$ & $m_{\rm DM}$ & $m_{\rm gas}$ & $\epsilon$ \\
   & $h^{-1} {\rm Mpc}$ & & $h^{-1} M_{\odot}$ & $h^{-1} M_{\odot}$ & $h^{-1}
   {\rm kpc}$ \\
  
  \hline
  
  D6 & 33.75 & $2 \times 486^3$ & $2.75 \times 10^7$ & $4.24 \times 10^6$ & 2.73 \\
  E6 & 50 & $2 \times 486^3$ & $7.85 \times 10^7$ & $1.21 \times 10^7$ & 4.12 \\

\hline

\multicolumn{6}{l}{$N_p$: Total number of particles} \\
\multicolumn{6}{l}{$m_{\rm DM}$: Mass of dark matter particles} \\
\multicolumn{6}{l}{$m_{\rm gas}$: Initial mass of gas particles} \\
\multicolumn{6}{l}{$\epsilon$: Comoving gravitational softening length} \\

\end{tabular}
\label{param}
\end{table}
Two simulation runs are analysed in this paper to allow for different volume
size and resolution. The main parameters are listed in Table \ref{param}.
Both runs were of moderate volume, with boxsizes of side length $33.75 h^{-1}
{\rm Mpc}$ (D6 simulation), and $50 h^{-1} {\rm Mpc}$ (E6). For both
simulations $N_p = $ $2 \times 486^3$ particles were used. The moderate
boxsizes prevent the simulations from being run below $z\sim 1$ to keep the
fundamental mode linear, but provide a large enough scale to produce
statistically significant quasar populations. The limitation on the boxsizes
is necessary to allow for appropriate resolution to carry out the subgrid
physics in a converged regime (for further details on the simulation methods,
parameters and convergence studies see \citet{DiMatteo2008}).

\begin{figure*}
  \centering

	\includegraphics[width=12cm]{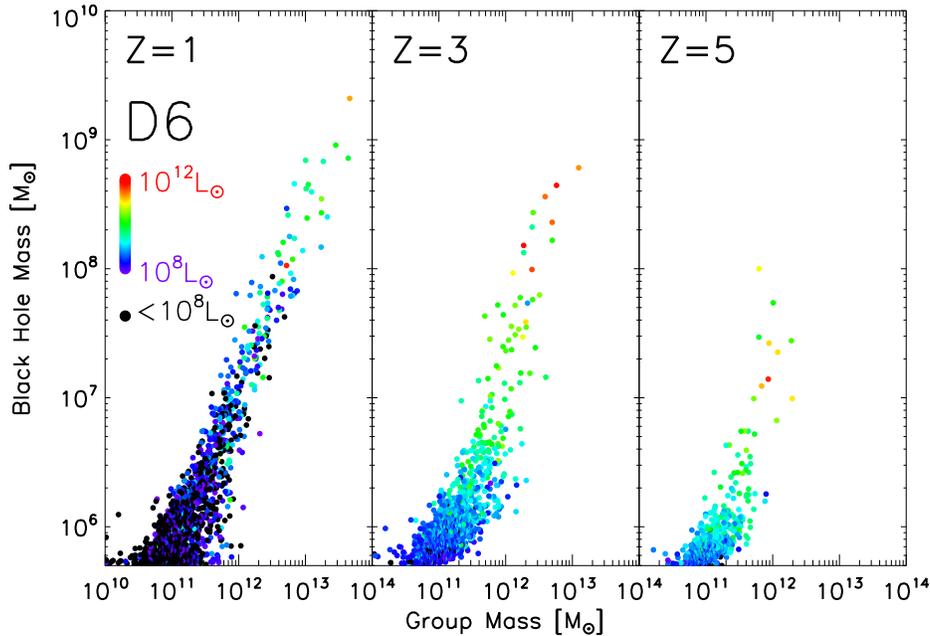}  \caption{Relation between masses of dark matter halos and their most massive
  black holes.  Color represents bolometric luminosity of
the massive BH.}
  \label{BHHaloRelation}
\end{figure*}

\subsection{Subgroup finder algorithm}
In addition to the on-the-fly friends-of-friends algorithm used to identify
groups, a modified version of the SUBFIND algorithm \citep{Springel2001} was
run on the FoF-identified groups to determine the component subgroups
(\textit{i.e.} galaxies) within each group.  These subgroups are defined as
locally overdense, self-bound particle groups.  To identify these regions, the
algorithm sorts the particles within the parent group by density, and then
analyzes each particle in order of decreasing density.  For each particle
\textit{i}, the density of the 32 nearest neighbors are checked.  If
none are denser than particle \textit{i}, it forms the basis for a new
subgroup.  If a single particle denser than \textit{i} is found, or if the
closest two denser particles belong to the same subgroup, particle \textit{i}
is assumed to be a member of that subgroup.  If the two nearest particles
denser than \textit{i} are members of different subgroups, these two subgroups
are stored as subgroup candidates, and are then joined into a new subgroup
also containing \textit{i}.  After checking each particle in this manner,
particles are checked for binding within their parent subgroup based on their
position relative to the position of the most bound particle, and the velocity
relative to the mean velocity of particles in the group.  Any particle with
positive total energy is considered unbound, and is removed from the subgroup,
leaving the group divided up into its component subgroups (galaxies).

\subsection{Correlation Function}

To investigate the clustering properties of quasars, we use the two-point
correlation function $\xi(r)$:
\begin{equation}
dP = \rho_0^2 [1+\xi (r)] dV_1 dV_2
\end{equation}
\citep{Peacock1999}, where dP is the probability of finding one object in each
volume element $dV_1$ and $dV_2$, separated by a distance r, with an average
number density of $\rho_0$.  We use the natural estimator $\xi(r) =
\frac{DD}{RR} -1$ for computing $\xi$, where DD and RR are the number of pairs
of objects found with separation r in the simulation (DD) and in a random
distribution of equal spatial density (RR).  For calculating RR, we used a
random distribution of $N_R = 6 \times 10^5$ objects to find the number of
pairs in a random sample, which is then normalized with a factor of $\left
  (\frac{N_{D}}{N_{R}} \right) ^2$ (where $N_D$ is the number of objects
considered for the DD term) to correct for the increased spatial density of
the random sources relative to the BHs in the DD term.  Note that the estimator $\xi(r) =
\frac{DD-2DR+RR}{RR}$ \citep{LandySzalay1993} has been shown to be more
accurate (as it more effectively accounts for edge effects), but when
considering small scales, both estimators provide equivalent results
\citep{Kerscher2000}.  Indeed, to confirm the validity of the natural
estimator, we compared results between the natural estimator and the Landy and
Szalay estimator, and found that for the largest scales ($> 5 \: \rm{h^{-1} \:
  Mpc}$) at low redshift, they differ by less than 5\%, and the discrepancy is
well below 1\% everywhere else.

\section{Results}
To illustrate the distribution of quasars (as a function of their luminosity)
with respect to the underlying matter distribution, in Figure
\ref{simulationslice} we plot a slice through the D6 simulation at $z=1,2,3,4$, with black hole positions indicated by colored dots for four
luminosity range bins: $L < 10^8 L_\odot$ - Orange; $10^8 L_\odot < L < 10^9
L_\odot$ - Pink; $10^9 L_\odot < L < 10^{10} L_\odot$ - Blue; $L > 10^{10}
L_\odot$ - Green.  As expected, as supermassive black holes are hosted by
galaxies, the quasars (particularly
the most luminous sources) are located in some of the densest regions, with low redshift tending to exhibit more BHs,
though at generally fainter luminosities.  To characterize the relation
between black hole and host halo mass more precisely, in Figure
\ref{BHHaloRelation} we show the relation between the group halo mass and the
mass of its most massive (central) black hole, with color representing the
respective (instantaneous) quasar luminosity.  There is a correlation between
halo mass and BH mass, and to a lesser extent between halo mass and BH
luminosity, with large halos tending to host more massive, more luminous black
holes than smaller halos, albeit with significant scatter. This is due to the
lightcurve that a black hole has in our simulations \citep[regulated by the
complex hydrodynamics, see e.g.][]{DiMatteo2008, DeGraf2010}.  We also find
that as redshift decreases, the simulation is more densely populated with BHs,
which tend to be more massive and less luminous than at earlier redshift.

\begin{figure}
  \centering
  \includegraphics[width=8cm]{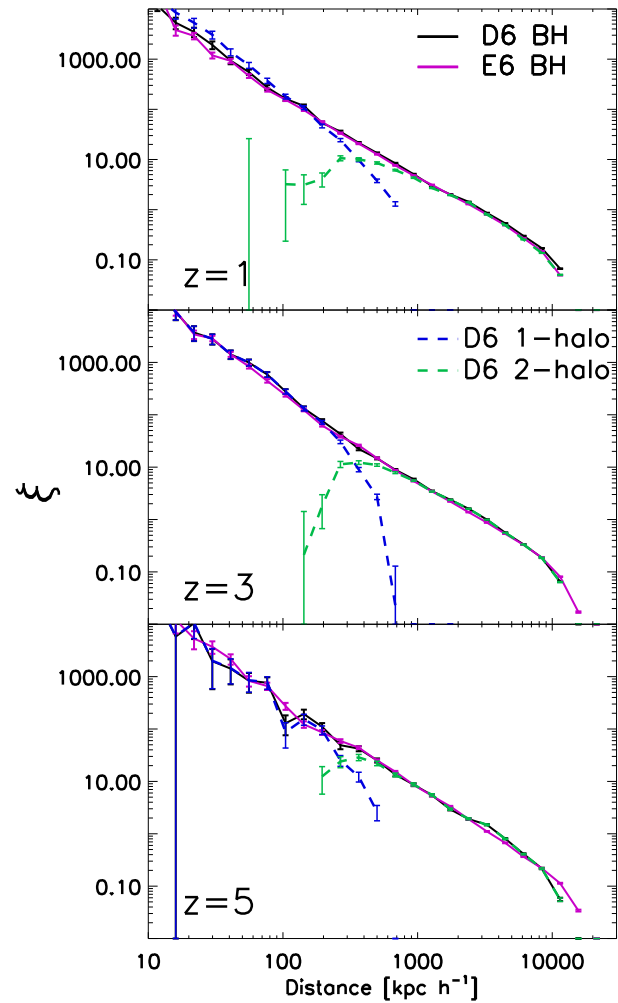}
  \caption{Two point correlation functions for the black holes in the D6
  (solid black) and E6 (solid pink) simulations at z=1, 3, 5, with the 1-halo
  and 2-halo terms for the D6 simulation explicitly shown (dashed blue and green, respectively).}
  \label{simcompare}
\end{figure}

To study the relation between black holes and other structures, in Figure
\ref{simcompare} we show the correlation functions of black holes found in the
D6 (solid black) and the E6 (solid pink) simulations for scales between $10 \:
\rm{kpc \: h^{-1}}$ and $\sim 10 \: \rm{Mpc \: h^{-1}}$ at z=1, 3, 5, with
Poisson error bars.  Note, the results from the two simulations are very
similar, with the higher resolution D6 simulation showing a small boost at
small scales (below $\sim 200 \: \rm{kpc \: h^{-1}}$).  In general, we see
$\xi_{BH}$ typically takes the form of a power law (with some possible excess
at small scales at $z=1$).

We also divide $\xi_{\rm{BH,D6}}$ into two terms: a 1-halo term (dashed blue)
produced by BH pairs occupying the same host group, and the 2-halo term
(dashed green) produced by pairs occupying different groups.  As expected,
the 2-halo term dominates at large scales (above $\sim 300 \: \rm{kpc \:
  h^{-1}}$), while at smaller scales the 1-halo term dominates, indicating
that our small scale clustering is really measuring BH properties within the
scales of the host halos.  A distinction between the 1-halo and 2-halo terms
is expected (as BHs are hosted by galaxies within halos) and is consistent
with the theoretical expectations \citep[see, e.g.][]{CooraySheth2002}, as
well as what has been found in galaxy correlation functions \citep[see,
e.g.][]{Magliocchetti2003, Zehavi2004}.

\subsection{Large Scale Clustering}
\begin{figure*}
  \centering
  \includegraphics[width=15cm]{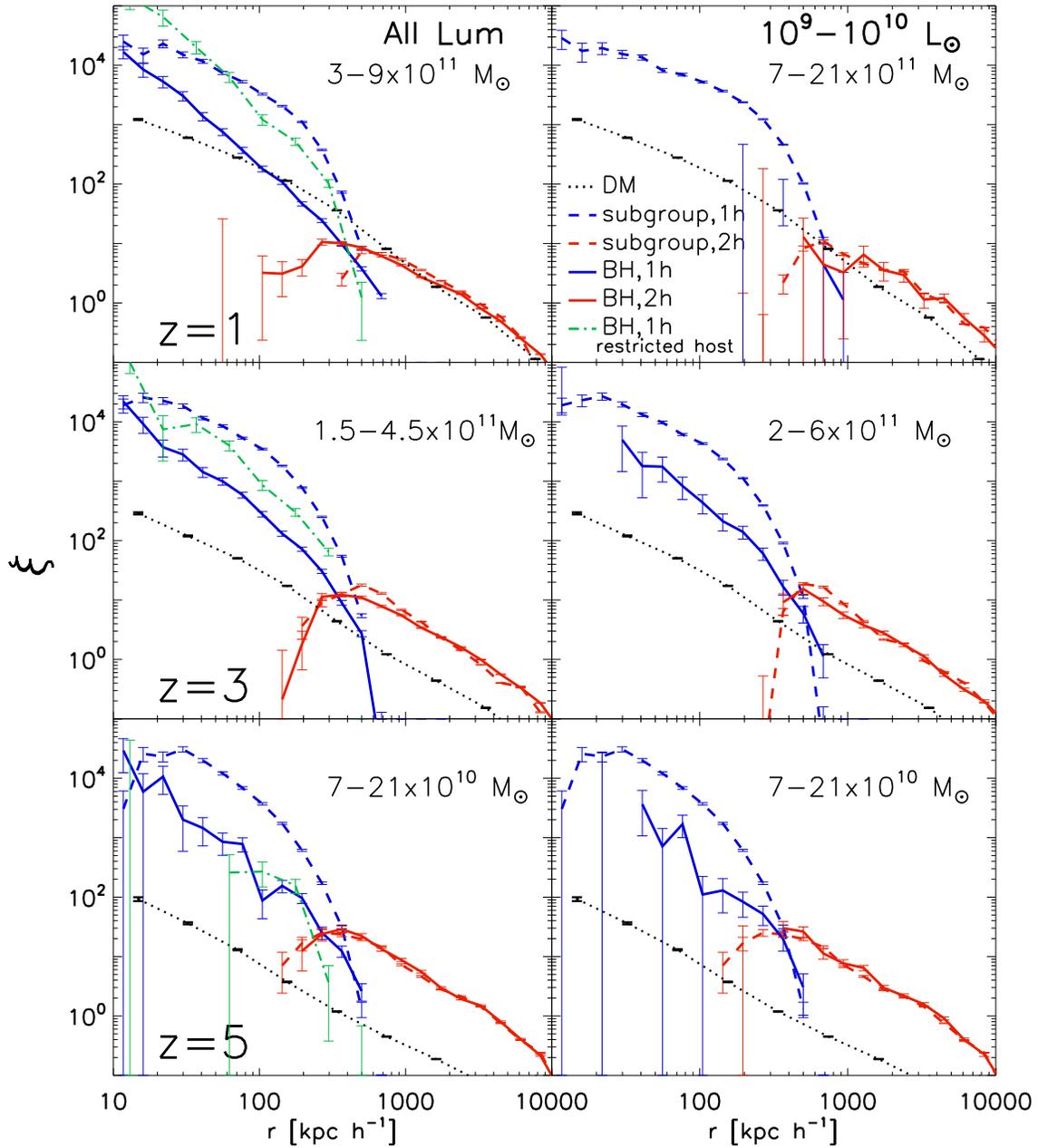}
      \caption{Correlation functions for the D6 simulation BHs (solid) and subgroups within a
      specified mass range (dashed), at z=1, 3, 5, with 1-halo and 2-halo terms
      plotted separately (blue and red, respectively).  The BH correlation function is plotted
      using all BHs (left) and using only those with $10^9 \: L_\odot <
      L_{\rm{BH}} < 10^{10} \: L_\odot$ (right).  The mass range for
      $\xi_{\rm{group}}$ is chosen so as to
      find the closest agreement between $\xi_{\rm{BH,2h}}$ and
      $\xi_{\rm{group}}$.  We also plot $\xi_{\rm{BH,1h}}$ using
      only BHs found in host groups in this fitted mass range (dot-dashed green line).}
      \label{typicalhalo}
\end{figure*}

It may be expected that black holes will cluster similarly to their host
galaxies (within their halos). To investigate the relation between BH
clustering and that of their host halos, in the left column of
Figure~\ref{typicalhalo} we plot the 1-halo (blue) and 2-halo (red)
contributions to the correlation function (at $z=1, \: 3, \: \rm{and} \: 5$)
for BHs (solid lines) and galaxies (as identified by the subgroup finder
described in Section 2.3) populating halos (\textit{i.e.} groups) in the
specified mass ranges (dashed lines).  These mass ranges were chosen to
reproduce the closest agreement between $\xi_{BH}$ and $\xi_{subgroup}$ in the
2-halo regime at each redshift, so as to be used as an indicator of the
typical halo mass for BH hosts (at each redshift).
The same is shown in the right column of Figure~\ref{typicalhalo} where we only
include some of the most luminous BHs in the simulations ($10^9 L_\odot <
L_{\rm{BH}} < 10^{10} L_\odot$, a range which is probed by observations).

For the full BH population, the typical host mass increases slightly with
decreasing redshift, from $\sim 10^{11} M_{\odot}$ to somewhat below $10^{12}
M_{\odot}$ from $z=5$ to $z=1$ respectively. When limited to the luminosity
range $10^9 L_\odot < L < 10^{10} L_\odot$, we again find increasing host mass
with decreasing redshift, but with a sharper increase up to masses a few times
$10^{12} M_{\odot}$ at $z=1$ \citep[still in the faint end of the luminosity
function, see][]{DeGraf2010}.

We compare the typical host mass found in this way to the mean (median) mass
of the host halos (see Table \ref{bhhosttable}) for several luminosity ranges,
and find that the 2-halo clustering as described above does indeed provide an
estimator for the mean host halo mass at the corresponding redshift in the
simulations. In addition, the table shows that for a given halo mass the
luminosity of its typical BH decreases with time, particularly at low redshift
(below $z \sim 2-3$), as seen more generally in Figure
\ref{BHHaloRelation}. This is shown explicitly in the bottom of Table
\ref{bhhosttable}, where we calculate the mean and median BH luminosities
found within groups of specified halo mass ranges.  Note that the mean quasar
luminosity actually peaks at $z = 3$ for massive ($M > 10^{12} M_\odot$)
groups as a result of a few highly luminous sources.

\begin{figure}
  \centering
    \includegraphics[width=8cm]{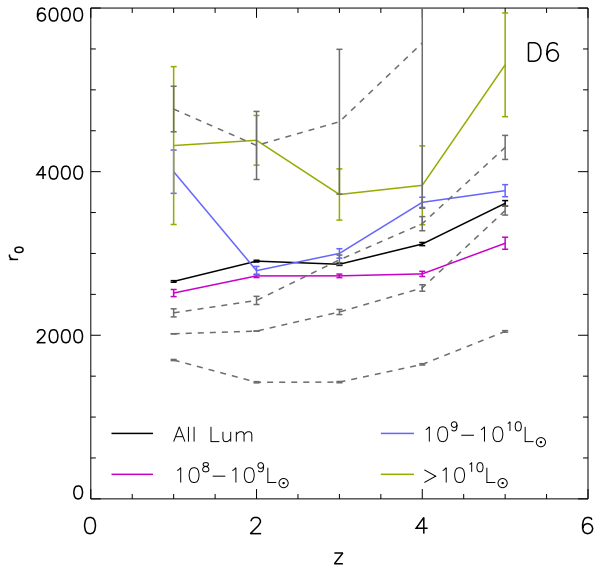}
\caption{\textit{Solid lines}: Black hole correlation length as a function of redshift for several luminosity bins (colored lines).
  \textit{Grey dashed lines}: Group correlation length as a function of redshift for group mass ranges (from top to bottom) $>10^{12} M_\odot$,
  $10^{11}-10^{12} M_\odot$, $5 \times 10^{10}-5 \times 10^{11} M_\odot$, $10^{10}-10^{11} M_\odot$.}
  \label{corrlen}
\end{figure}

To better characterize the overall clustering strength, and in particular its
luminosity dependence and evolution with redshift, we use the correlation
length $r_0$, defined as the scale at which $\xi(r_0) = 1$ [which we calculate
using a linear extrapolation of $\xi$]. In Figure \ref{corrlen} we plot $r_0$
versus $z$ for BHs in several luminosity bins (solid colored lines) and, for
comparison, groups in several mass bins (dashed grey lines). In general we
find a weak evolution of the quasar clustering with redshift. This can be
simply explained by the evolution of the bias of its underlying host halo
masses.  In particular, the correlation length for luminous ($L > 10^9
L_\odot$) BHs tends to decrease slightly as a function of decreasing redshift
until $z \sim 3$, following closely the bias of the $10^{11}-10^{12} M_\odot$
groups (consistent with the constraints on the host masses of these BHs).  At
fixed mass, these groups are less biased as a function of decreasing redshift
\citep{MoWhite2002,Bahcall2004}.  This is also in accord with our results from
Figure \ref{typicalhalo}, that the typical host halo mass remains roughly
constant for $z > 3$. For lower redshift (particularly $z < 2$), we instead
see a significant upturn in $r_0$ versus $z$, corresponding to the increase in
typical host halo mass, just as we found in Figure \ref{typicalhalo} and Table
\ref{bhhosttable}.  The lowest luminosity sources, however, show only minor
change in $r_0$, corresponding to a host mass which changes only slightly with
redshift (consistent with the median host masses found in Table
\ref{bhhosttable}).

This luminosity dependence is sufficiently weak (less than a factor of 2
increase in $r_0$ across several orders of magnitude in luminosity) to remain
broadly consistent with the predictions from models that suggesting bright and faint
quasars occupy similar halos \citep[e.g.][]{Lidz2006, Bonoli2009}.  Indeed, our simulations
produce complex lightcurves for our black holes, with luminosity varying
rapidly across several orders of magnitude \citep[see, e.g.][]{DiMatteo2008,
  DeGraf2010}.  This produces significant scatter in the relation between
black hole luminosity and host mass, so general agreement with
lightcurve-based models is expected (which are indeed motivated by simulations
similar to our own).  Nonetheless, as seen in Figure \ref{BHHaloRelation},
there remains some correlation between BH instantaneous luminosity and group
mass, so a weak dependence on luminosity is expected even in this model.

\subsection{Small Scale Clustering}
Although we find that the 2-halo terms for BHs and subgroups (galaxies) can be
easily matched to provide a good estimator for typical host mass, there is
significant discrepancy between their respective 1-halo terms
(Figure \ref{typicalhalo}, blue lines). The 1-halo BH correlation function is
different both in shape and amplitude to the 1-halo term of galaxies,
suggesting that, unlike at large scales, BHs do not cluster like their host
galaxies on small scales. Or in other words, the distribution of BHs within
halos does not follow closely that of their galaxies and hence does not trace
the underlying matter distribution.

In terms of amplitude, $\xi_{\rm{BH,1h}}$ can be adjusted by only considering
the BHs in those groups that match the mass range constrained by the 2-halo
term, thereby minimizing the suppression of $\xi_{\rm{BH,1h}}$ from the
numerous BHs in groups too small to contribute to the 1-halo term (due to
hosting only a single BH). As expected, in this case, (Figure \ref{typicalhalo},
green line) the amplitude increases and is more in agreement with the 1-halo
term of the subgroups ($\xi_{\rm{subgroup,1h}}$; at least at z=1-3 where the
statistics are good enough).

\begin{figure}
\centering
\includegraphics[width=8cm]{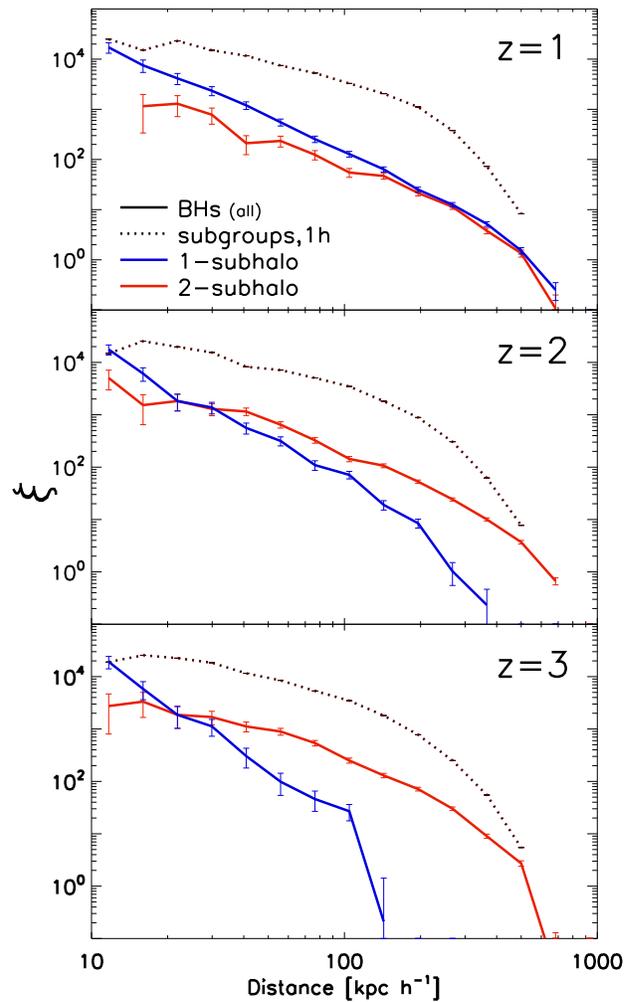}
\caption{\textit{Solid lines}: The 1-halo BH correlation function at z=1,2,3
  divided into components from BH pairs
  occupying separate subhalos (red) or co-habitating a single subhalo (blue), using the
  full population of BHs.  \textit{Dotted lines}: The 1-halo subgroup
  correlation function at z=1,2,3.}
\label{1halo}
\end{figure}

It is, however, hard to account for the substantial difference in shape:
$\xi_{\rm{BH,1h}}$ follows an approximate power law, lacking the decrease in
slope at small scales (below $\sim 200-300 \: \rm{kpc} \: h^{-1}$) observed in
$\xi_{\rm{subgroup,1h}}$ and expected from the 1-halo clustering produced by a
general NFW profile \citep{NFW1996, CooraySheth2002, Zehavi2004}.  Thus the
BHs are distributed significantly differently than an NFW profile, showing a
significant boost at small scales. 

We investigate the reason for this difference in the shape of the BH 1-halo
term in terms of multiple BHs co-existing in a given subgroup. These BHs end
up in a given subgroup as a result of mergers between their host galaxies, 
so that multiple BHs are expected to co-exist in a remnant (until dynamical
friction is able to bring them close enough together to eventually merge).

To understand the effect this has on the small scale clustering of BHs, we
calculate the contributions to $\xi_{\rm{BH,1h}}$ from pairs of BHs occupying
the same galaxy (we will call this the '1-subhalo' term) and from pairs of BHs
occupying different galaxies within the same group ('2-subhalo' term), in
analogy with dividing the overall correlation function into its 1-halo and
2-halo terms. We note that the existence of multiple BHs within a single
subgroup necessarily indicates a previous merger event, since BH particles are
not inserted into galaxies which already contain a BH particle, and thus any
1-subhalo contribution is inherently a result of previous galaxy mergers.

\begin{figure}
\centering
\includegraphics[width=8cm]{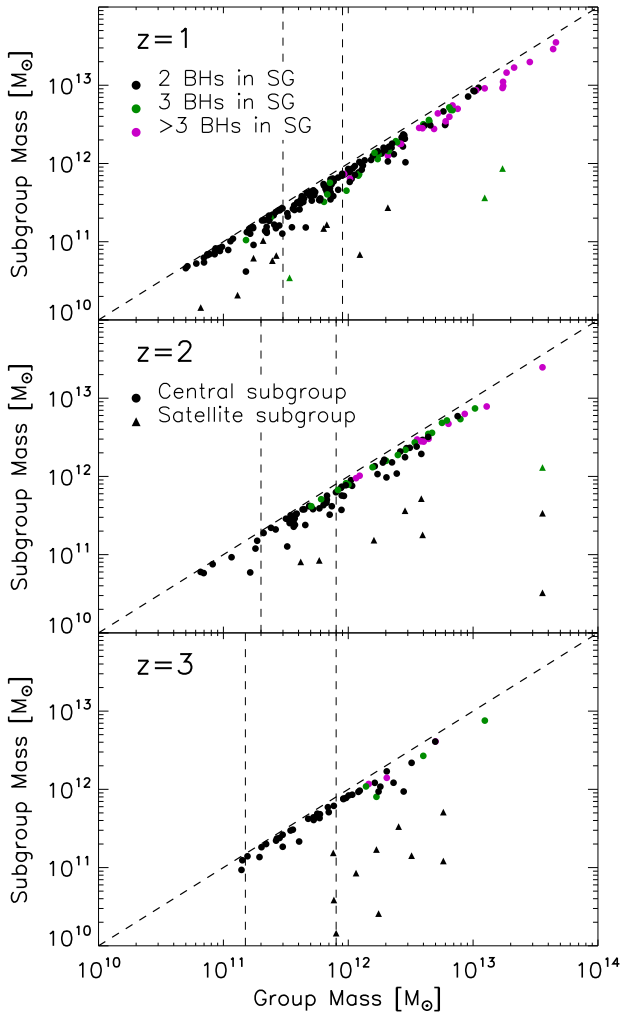}
\caption{The mass of each subgroup containing at least 2 BHs vs. the mass of
  its host group.  Color indicates number of BHs within the given subgroup: black - 2 black holes; green - 3 black holes; blue - 4 black
  holes; pink - more than 4 black holes.  Symbol indicates if the subgroup is
  the primary (\textit{i.e.} central) subgroup (circle), or a satellite
  subgroup (triangle).  \textit{Dotted line}: Represents a
  one-to-one mass ratio provided for reference.}
\label{groupvssg}
\end{figure}

\begin{figure}
\centering
\includegraphics[width=8cm]{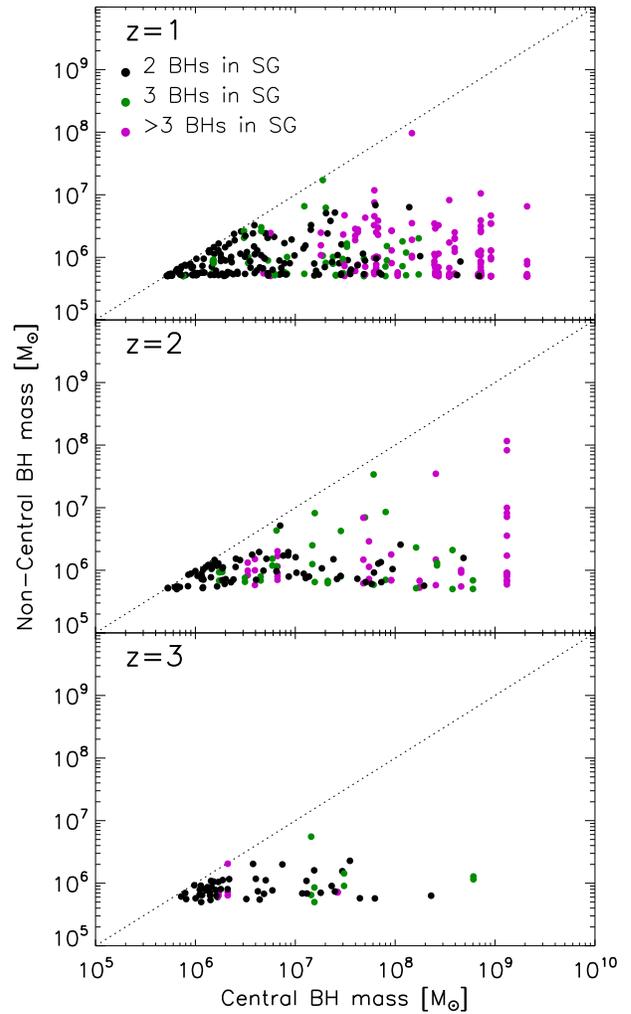}
\caption{The mass of the central (largest) BH within a multiply-occupied
  subgroup relative to the mass of the non-central BHs within the same
  subgroup.  Color indicates the number of BHs contained within the given
  subgroup: black - 2 black holes; green - 3 black holes; blue - 4 black
  holes; pink - more than 4 black holes.  \textit{Dotted line}: Represents a
  one-to-one mass ratio provided for reference.}
\label{relativemass}
\end{figure}

In Figure \ref{1halo} we plot the 1-subhalo (solid blue) and 2-subhalo (solid
red) components of $\xi_{\rm{BH,1h}}$, together with $\xi_{\rm{subgroup, 1h}}$
for subgroups in groups within the mass ranges listed in Figure
\ref{typicalhalo} (dotted line). The 1-subhalo term does indeed have a steeper
slope than the 2-subhalo term, and is most significant at small scales. The
1-subhalo term is most dominant at low redshift, and by $z=1$ it dominates the
entire 1-halo term. This is a result of having increasingly large groups at
low redshift which have also undergone a relatively large number of mergers.
These indeed contain multiply-occupied subgroups (see Figure \ref{groupvssg}).
We also find that, if restricted to BHs within the same host mass range as the
subgroups, the 2-subhalo term of $\xi_{\rm{BH,1h}}$ matches
$\xi_{\rm{subgroup,1h}}$ quite closely.  Thus we find that, within
sufficiently large groups (such that the simulation contains enough groups
hosting multiple BHs to produce a well-defined 1-halo term),
$\xi_{\rm{BH,1h}}$ has two distinct components: one due to BH pairs which
occupy separate galaxies, exhibiting good agreement with
$\xi_{\rm{subgroup,1h}}$; and a steeper one caused by BH pairs which co-occupy
individual galaxies as a result of previous galaxy mergers, causing a boost in
the small-scale $\xi_{\rm{BH,1h}}$, particularly evident at low redshift,
where typical groups are largest and have undergone significant merging.

In Figure
\ref{groupvssg} we plot the relative mass of each multiply-occupied subgroup
and its host group, with circles indicating central subgroups, and triangles
showing satellite subgroups.  We clearly see that these multiply-occupied subgroups tend to be the primary
(central) subgroup within a given group, typically containing $\sim 65-70\%$ of the total group's mass.  We also
color-code the datapoints to show the number of BHs within each subgroup, and
see that the central subgroup of larger groups tends to contain more BHs.

To investigate the masses of BHs which populate these multiply-occupied
subgroups, in Figure \ref{relativemass} we plot the mass of the largest
(primary) black hole within a given subgroup relative to the masses of the
other BHs in the same subgroup, color-coded to show the number of black holes
within the subgroup.  In only a few rare cases do we have more than one
massive BH, while in the majority of cases we have, at most, a single massive
BH with one or more smaller black holes, generally within an order of
magnitude of the seed mass.  This suggests that the majority of BHs in
multiply-occupied subgroups come from relatively small satellite subgroups
(hosting correspondingly small black holes) which have fallen in and merged
with the large, central subgroup, but do not grow substantially, instead
remaining much less massive than the primary BH in the galaxy.  Additionally,
we observe that over time the fraction of BHs in the simulation located within
these multiply-occupied subgroups increases from 2\% at $z=5$ to 15\% at
$z=1$, as typical groups get larger and have had more opportunity for
satellite subgroups to merge with the central subgroup. This increase in
typical group mass causes an increase in both the number of multiply-occupied
subgroups, as well as an increase in the typical number of black holes found
within them (as seen in Figure \ref{relativemass}), which produces the
increased importance of the 1-subhalo term with decreasing redshift seen in
Figure \ref{1halo}.

We will compare the small scale clustering from the simulations to
observations in Section 3.4.

\begin{table*}
\caption{Mean (median) halo mass of parent group and mean (median) luminosity
  of daughter BHs in D6 simulation. }
\begin{tabular}{c c c c c c}

  \hline
  \hline
  
  & z=1 & z=2 & z=3 & z=4 & z=5 \\
\hline
  BH Luminosity & \multicolumn{5}{c}{Mean (Median) Group Mass [$10^{10} M_\odot$]} \\

    All &  39.9 (10.1) & 24.2(9.07) & 19.3 (9.51) & 14.8 (8.34)& 12.6 (8.34)\\
$10^8 L_\odot< L_{\rm{BH}} < 10^9 L_\odot$  & 61.5 (18.7) & 27.2 (9.66) & 18.9
    (8.80) & 13.1
    (7.53) & 11.5 (7.12)\\
$10^9 L_\odot < L_{\rm{BH}} < 10^{10} L_\odot$ & 252 (94.1) & 54.9 (20.4) & 38.2 (21.1)
    & 25.5 (14.9) & 15.9 (10.3)     \\

  \hline


\hline
  Group Mass & \multicolumn{5}{c}{Log(Mean (Median) BH Luminosity) [log($L_\odot$)]} \\

  $M_{\rm{group}} < 10^{11} M_\odot$ & 7.88 (7.69) & 8.96 (8.55)& 8.99 (8.83)&
  9.39 (9.25)& 9.36 (9.34)\\
  $10^{11}M_\odot < M_{\rm{group}} < 10^{11.5} M_\odot$ & 8.66 (7.86)& 9.09
  (8.76)& 9.39 (9.07)&
  9.54 (9.31)& 9.71 (9.48)\\
  $10^{11.5}M_\odot < M_{\rm{group}} < 10^{12} M_\odot$ & 9.09 (8.23)& 9.64
  (9.19)& 9.90 (9.42)&
  10.19 (9.64)& 10.65 (9.88)\\
  $10^{12}M_\odot < M_{\rm{group}} < 10^{12.5} M_\odot$ & 9.45 (8.85)& 10.19 (9.81)&
  11.16 (10.34)& 10.82 (10.66)& 10.92 (10.68)\\
  $ 10^{12.5} M_\odot < M_{\rm{group}}$ & 10.49 (9.51)& 10.33 (10.20)& 12.40
  (11.49)& 11.48 (11.54)& N/A\\

  \hline

\end{tabular}
\label{bhhosttable}
\end{table*}

\subsection{Quasar Bias}
To further characterize the clustering properties of BHs, we now consider the
quasar bias as a function of scale and redshift. The bias is obtained by
taking the square root of the ratio between $\xi_{\rm{BH}}$ and the DM
correlation function (shown as dotted lines in Figure \ref{typicalhalo}).
Based on our results of the small scale clustering, we expect the quasars to
be strongly biased with respect to the DM distribution at small scales,
particularly at high redshift. This general trend is clearly seen in Figure
\ref{typicalhalo}, where $\xi_{\rm{DM}}$ (dotted lines) increases with
time (due to gravitational collapse), while $\xi_{\rm{BH}}$ tends to decrease
slightly (seen more clearly in Figure \ref{corrlen}).  More importantly, we
see that the BH clustering bias relative to that of DM is strongly
scale-dependent, with $\xi_{\rm{BH}}$ exhibiting a significant increase in
clustering at small scales (below $\sim 300 \: \rm{kpc \: h^{-1}}$) due to the
strong 1-halo term, whereas $\xi_{\rm{DM}}$ shows only a slight increase at
these small scales.

\begin{figure}
  \centering
  \includegraphics[width=8cm]{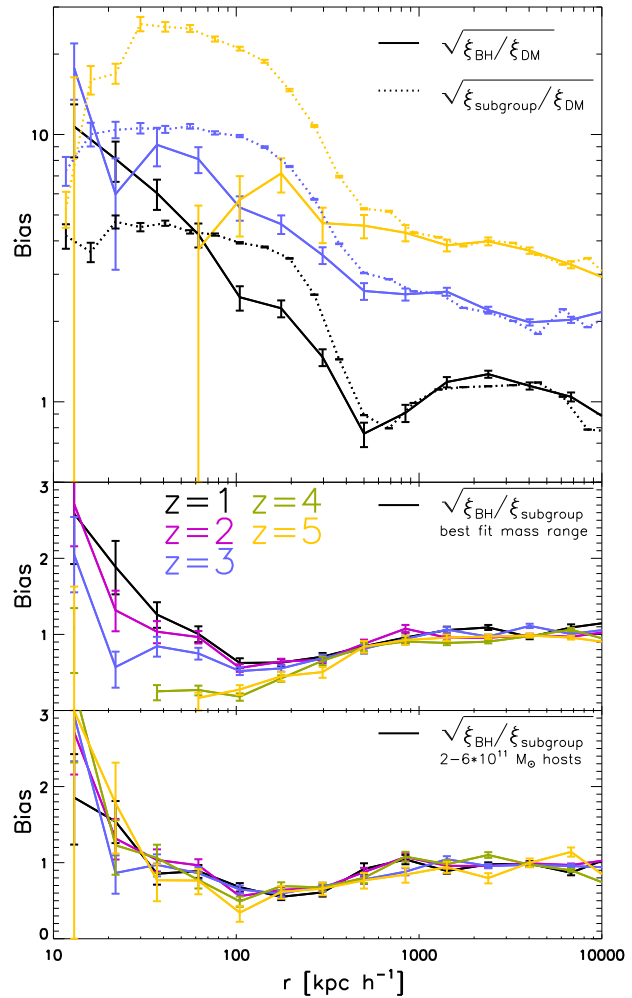}
  \caption{\textit{Top:} Solid lines: Black hole bias defined as
  $\sqrt{\xi_{\rm BH}/\xi_{\rm DM}}$, using only
  BHs occupying halos in the best-fitting mass ranges specified in Figure
  \ref{typicalhalo}.  Dotted lines: Subgroup bias
  defined as $\sqrt{\xi_{\rm subgroup}/\xi_{\rm DM}}$, using only
  subgroups occupying halos in the best-fitting mass ranges specified in Figure
  \ref{typicalhalo}.  \textit{Middle:} Bias of
  BHs relative to subgroups $\left (\sqrt{\xi_{\rm
  BH}/\xi_{\rm subgroup}} \right )$ occupying halos in the typical mass ranges
  found in Figure \ref{typicalhalo}. \textit{Bottom:} Bias of
  BHs relative to subgroups $\left (\sqrt{\xi_{\rm
  BH}/\xi_{\rm subgroup}} \right )$ occupying halos of mass $2-6 \times
  10^{11} \: M_\odot$.
}
  \label{bias}
\end{figure}

In the top of Figure \ref{bias} we plot the scale-dependent BH bias and
subgroup bias (defined as $\sqrt{\xi_{\rm BH}/\xi_{\rm DM}} \: ;
\sqrt{\xi_{\rm subgroup}/\xi_{\rm DM}}, \: \rm{respectively}$) found within
the hosts of the best-fitting mass ranges found in Figure \ref{typicalhalo}.
Here we see that the subgroup bias levels off (as did $\xi_{\rm{subgroup}}$ in
Figures \ref{typicalhalo} and \ref{1halo}), but the 1-subhalo term causes the
BH bias to continue increasing to the smallest scales probed in our
simulation.  To show this more clearly, the middle of Figure \ref{bias} shows
the bias of BHs relative to the subgroups $\left (\sqrt{\xi_{\rm
      BH}/\xi_{\rm subgroup}} \right )$ for z=1-5. Within a given host mass
range, the BHs cluster very similarly to the subgroups (galaxies), except at
the smallest scales (below $\sim 100 \: \rm{kpc \: h^{-1}}$), where we again
see the increased clustering caused by the multiply-occupied subgroups
remaining from merger events, as discussed earlier.  Although we note that
this small-scale bias appears to be redshift dependent, it is actually a
result of the evolution of the host mass being considered.  At higher
redshifts, the typical host mass is smaller, and thus fewer will have
undergone subgroup mergers producing multiply-occupied subgroups (as confirmed
in Figure \ref{groupvssg}), thereby making the small-scale boost less
apparent.  When considering behavior for a fixed group mass (as shown in the
bottom of Figure \ref{bias}), we see that the bias between BH and subgroup
clustering is redshift-independent, and consistently exhibits a strong
small-scale boost from past subgroup mergers.

\begin{figure*}
  \centering
\includegraphics[width=18cm]{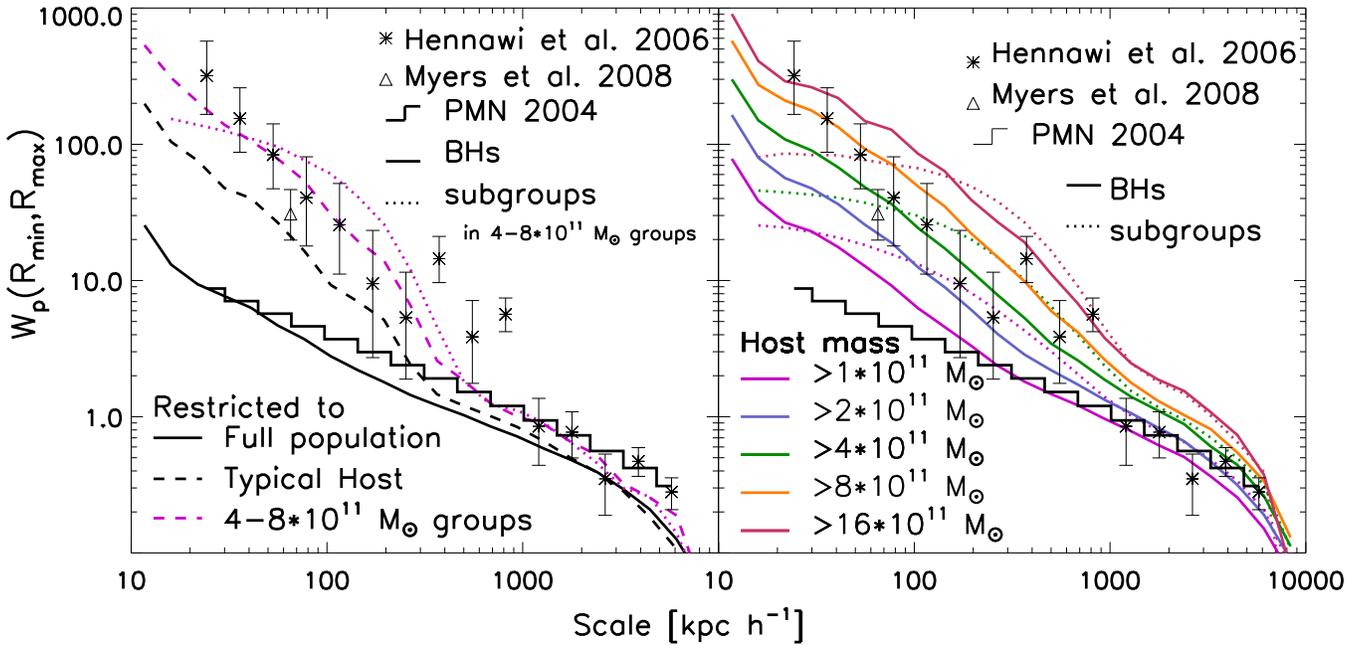}

  \caption{\textit{Left:} The projected correlation function from the D6 simulation, averaged
  across redshifts 1-3, and across 3 projected directions for the full BH
  population (solid line), for BHs found within groups of the typical host
  mass shown in Figure \ref{typicalhalo} (dashed black line), for BHs found in
  groups of mass $4-8 \times 10^{11} M_\odot$ (dashed pink line), and for
  subgroups found in groups of mass $4-8 \times 10^{11} M_\odot$ (dotted pink line).  We also plot
  the extension of the power law found in \citet{Porciani2004} (step
  function), and the observational results of \citet{Hennawi2006} (asterisks)
  and \citet{Myers2008} (triangle).  \textit{Right:} Same as left plot, but
  with $\overline{W}_P (R_{\rm{min}}, R_{\rm{max}})$ plotted for several
  lower-limits on the host group mass.}
  \label{projectedxi}
\end{figure*}

\subsection{Comparison with observations:  Projected Correlation Function}

In order to compare with the observational constraints on the small scale
clustering \citep[see][]{Hennawi2006, Myers2008}, we compute the
volume-averaged projected correlation function $\overline{W}_P (R_{\rm{min}},
R_{\rm{max}})$.  This projected correlation function is computed using the
same estimator described in Section 2.4, but the separation between points is
the projected separation onto the \textit{xy}, \textit{xz}, or \textit{yz}
plane, rather than the separation in three-space.  Although these three
projections provide comparable results, we average across the three directions
to avoid any potential directional bias.  We then average across redshifts 1-3
(to match the observational data redshift range), weighted by the number of
BHs at each redshift, and plot the result in Figure \ref{projectedxi},
together with the data from \citet{Hennawi2006} (asterisks), \citet{Myers2008}
(triangle), and the extension of the best-fit power law for the large scale
clustering found by \citet{Porciani2004} (step-function).  We have also
plotted the projected correlation function for subhalos found in our
simulations for several host mass ranges (dotted lines).

Figure \ref{projectedxi} shows a remarkable agreement between the small scale
clustering of BHs from the simulations with the observations.  In particular,
when considering BHs within groups in the mass range of $4-8 \times 10^{11}
M_{\odot}$ the small scale boost (magenta line) matches the observed clustering
very well. For completeness we also show (dashed black line) the signal
expected from BHs in the hosts of the typical mass ranges shown in Figure
\ref{typicalhalo} which is also in good agreement, although slightly lower
normalization. Indeed, the observed small-scale excess can be explained as
resulting from the merger-based boost found in our simulations, further
emphasizing the importance of such mergers on quasar evolution.

We also note that if the full BH population from our simulation is
used, rather than those in the restricted mass range, we lose the small scale
excess (solid line), since the majority of our BHs are found in groups too
small to exhibit significant effects of subgroup mergers.

To investigate the dependence of the projected correlation function on the
host mass in more detail, in the right of Figure \ref{projectedxi} we plot
$\overline{W}_P (R_{\rm{min}}, R_{\rm{max}})$ for BHs hosted by groups with
several different lower-mass cutoffs (from $1 \: \rm{to} \: 16 \times 10^{11}
M_\odot$; colored lines), together with the observational data.  Here we see
that including less massive groups causes an overall decrease in amplitude (as
expected), and also suppresses the small scale excess, as a result of smaller
groups being less likely to host a multiply-occupied subgroup.  This suggests
that, given sufficient observational data, small-scale clustering may provide
a sensitive means of probing the typical mass of merging pairs of galaxies
hosting supermassive black holes.  As shown, the curves with a lower
mass cut of $\sim 4-8 \times 10^{11} M_\odot$ produce the best agreement with
observation, implying that observed quasar pairs are typically located within
groups of moderate size, comparable to those found within our simulation (and
thus below the larger host masses typically associated with observed
large-scale clustering).

\section{Conclusions}
In this paper we have investigated the clustering of black holes
within hydrodynamic cosmological simulations, its redshift evolution,
luminosity dependence, and particularly the small-scale behavior. 

We have shown that the large scale clustering of black holes traces that of
the galaxies within their host groups, and provides a predictor of the typical
host mass, which for our simulations is found to be on the order of a few
$10^{11} M_\odot$.  Although well below the typically found masses of $\sim 2
\times 10^{12}-10^{13} M_\odot$ \citep{Lidz2006,Ross2009,Bonoli2009,
  Shen2009}, this is consistent with our limited simulation volumes which can
only follow the growth of the faint-end of quasar population (DeGraf et
al. 2010), and cannot follow formation of such massive groups. The typical
host group mass shows some evolution with redshift, most significant below $z
\sim 3$, where typical host masses increase by up to a factor 10 (at $z=1$).
This low-redshift increase is distinctly luminosity dependent, with the more
luminous sources ($L_{\rm{BH}} > 10^9 L_\odot$) undergoing the most
substantial increase in typical host mass.  Overall the evolution of
clustering with redshift and luminosity is minor and consistent with current
observational constraints (albeit in low luminosity populations this is yet to
be fully constrained). The relatively weak dependence found in our
simulations is consistent with the complex lightcurves we derive from our
direct modeling in which quasar luminosities vary over relatively short
timescales for a given host (as regulated by hydrodynamical processes).  This
is also consistent with the models of \citet{Lidz2006}.

In addition to the large-scale clustering (the 2-halo regime), our simulations
allow us to study the small scale clustering (the 1-halo term) of
$\xi_{\rm{BH}}$.  We found that $\xi_{\rm{BH,1h}}$ follows a power law
behavior all the way to the smallest scales. The clustering of black holes at
small scale is unlike that of galaxies (or dark matter).  We showed that
the 1-halo BH term can be subdivided into two components: 1-subhalo and
2-subhalo. The 1-subhalo term, $\xi_{\rm{subgroup,1h}}$, represents the
clustering of BHs within a galaxy and 2-subhalo that of BHs occupying
different galaxies.  We have shown that the 1-subhalo is the one that provides
the power law behavior, indicating that galaxies do contain multiple black
holes as a result of mergers. These galaxies tend to be the central galaxy
within relatively large groups (for our simulation), generally hosting at most
a single massive BH with one or more smaller BHs, likely as a result of
smaller satellite galaxies merging with the large, central galaxy within the
group.  In the absence of these multiply-occupied galaxies, $\xi_{\rm{BH,1h}}$
and $\xi_{\rm{subgroup,1h}}$ exhibit very close agreement, but the inclusion
of these merger remnants causes a significant boost in the small-scale BH
clustering.  This merger-based boost is most significant at low redshift,
where typical group size is largest, though we find it in sufficiently massive
groups at all redshifts.

Though observational limitations make observing these scales difficult,
several recent studies have found a small scale excess at scales below $\sim
100 \: \rm{kpc} \: h^{-1}$ \citep{Hennawi2006, Myers2008}. The observed excess
is in remarkable agreement to the one predicted by our simulations coming from
groups approaching $10^{12} M_\odot$, which host mostly intermediate size
black holes. This suggests that multiple black holes co-occupying a subgroup at
low redshifts are likely faint(ish) AGNs hosted in Milky Way size halos that
have recently undergone merging.  We also note that galaxies hosting multiple
AGN \citep{Komossa2003, Gerke2007, Barth2008, Comerford2009b} or inspiralling
supermassive black holes \citep{Comerford2009} have been found in recent
studies, further supporting our conclusion of multiply-occupied subgroups.
Although we leave more detailed investigation of the small scale BH pairs in
our simulations (particularly with regard to the luminosities of inspiralling
black holes) for a future work, we note that our finding that
multiply-occupied galaxies tend to host a single massive BH with one or more
small BHs appears to be in keeping with the observation that most of the
inspiralling BH pairs power only a single AGN \citep{Comerford2009}.  Given
that, our agreement in small-scale merger-induced boost certainly reinforces
the importance of galaxy mergers on the evolution of supermassive black holes.
We also note this small-scale excess' sensitivity to the host mass suggests
that future small-scale studies may provide a means to constrain the typical
mass of merger events between galaxies hosting black holes, with current
observational data combined with our simulations suggesting groups with
typical masses comparable to those probed in our simulations (from a few
$10^{11} M_\odot$ to $10^{12} M_\odot$) produce the multiply-occupied galaxies
underlying the observed small scale excess.

We would like to point out however that there are several aspect of our
modeling approach, including numerical issues, in the simulations that
potentially affect our results on the small-scale clustering.  We have a very
simplistic prescription to determine how BHs merge with one another (imposed
by the limits on the resolution that can be achieved in these cosmological
boxes). The current prescription has a BH pair merge when BHs are separated by
less than their smoothing length and if the BHs relative velocity is small
compared to the local sound speed. Changes to this prescription could
accelerate (postpone) BH mergers, which would result in a suppression
(increase) of our small scale clustering signals. It would be desirable to
compare our results with other simulations which implement different
prescriptions, or in the future to include more direct physical modeling of
this region in higher resolution simulations.  However, neither of these are currently possible.  A numerical issue that may affect the results of our
one-halo term is that black holes need to be fixed to potential minima
(calculated among the neighboring particles within the smoothing length used
for the accretion model) in order to avoid them leaving their subhalo due to
numerical N-body noise (and the fact that dynamical friction is hard to
calculate for sink particles).  However, in some instances this may cause a BH
particle in a small subhalo in orbit in a bigger group to 'hop' to
the potential minimum of the larger group. This effect may be exacerbated in
situations where the small subhalo may be stripped of gas by infalling into a
larger one. These effects could artificially increase the number of BHs within
large, central halos, thereby boosting small scale clustering. However, when
we measure what fraction of BHs appear to 'hop' into the center of groups
experiencing an unexpected jump in their position, we find that it is only
$\sim 1-2\%$.  Future simulations and comparison amongst different approaches
(once they become available) should of course attempt to characterize these
effects more specifically.  We further note however, that observational
studies have indeed found cases of galaxies hosting multiple BHs
\citep{Comerford2009}, so the existence of a one-subhalo term is expected.
Additionally, as seen in Figure \ref{projectedxi}, the projected clustering of
subgroups has a fundamentally different form than the observed quasar
clustering.  Thus the BHs cannot simply trace their host subgroups/galaxies
and still produce the observed small scale excess, but rather a significant
one-subhalo term is required to produce the small scale power law behavior.

In future work we also plan to simulate larger volumes (which we are starting
to be feasible with the most advanced technology) to allow us to study
clustering of AGN at larger (mass and length) scales while simultaneously
investigating luminosity dependence for brighter sources more directly
comparable to current and upcoming observational data, as well as providing
increased statistics for the small scale clustering.

\section*{Acknowledgments}

This work was supported by the National Science Foundation, NSF Petapps,
OCI-079212 and NSF AST-0607819.  The simulations were carried out at the NSF
Teragrid Pittsburgh Supercomputing Center (PSC).

 \bibliographystyle{mn2e}	
 \bibliography{astrobibl}	

\begin{thebibliography}{60}
\expandafter\ifx\csname natexlab\endcsname\relax\def\natexlab#1{#1}\fi

\bibitem[{Bahcall} et~al.(2004){Bahcall}, {Hao}, {Bode} \& {Dong}]{Bahcall2004}
{Bahcall} N.~A., {Hao} L., {Bode} P., {Dong} F., 2004, \apj, 603, 1

\bibitem[{Barth} et~al.(2008){Barth}, {Bentz}, {Greene} \& {Ho}]{Barth2008}
{Barth} A.~J., {Bentz} M.~C., {Greene} J.~E., {Ho} L.~C., 2008, \apjl, 683,
  L119

\bibitem[{Bondi}(1952)]{1952MNRAS.112..195B}
{Bondi} H., 1952, \mnras, 112, 195

\bibitem[{Bondi} \& {Hoyle}(1944)]{1944MNRAS.104..273B}
{Bondi} H., {Hoyle} F., 1944, \mnras, 104, 273

\bibitem[{Bonoli} et~al.(2009){Bonoli}, {Marulli}, {Springel}, {White},
  {Branchini} \& {Moscardini}]{Bonoli2009}
{Bonoli} S., {Marulli} F., {Springel} V., {White} S.~D.~M., {Branchini} E.,
  {Moscardini} L., 2009, \mnras,  606

\bibitem[{Comerford} et~al.(2009{\natexlab{a}}){Comerford}, {Gerke}, {Newman}
  et~al.]{Comerford2009}
{Comerford} J.~M., {Gerke} B.~F., {Newman} J.~A., et~al., 2009{\natexlab{a}},
  \apj, 698, 956

\bibitem[{Comerford} et~al.(2009{\natexlab{b}}){Comerford}, {Griffith}, {Gerke}
  et~al.]{Comerford2009b}
{Comerford} J.~M., {Griffith} R.~L., {Gerke} B.~F., et~al., 2009{\natexlab{b}},
  \apjl, 702, L82

\bibitem[{Cooray} \& {Sheth}(2002)]{CooraySheth2002}
{Cooray} A., {Sheth} R., 2002, Phys. Rep., 372, 1

\bibitem[{Croom} et~al.(2005){Croom}, {Boyle}, {Shanks} et~al.]{Croom2005}
{Croom} S.~M., {Boyle} B.~J., {Shanks} T., et~al., 2005, \mnras, 356, 415

\bibitem[{Croom} \& {Shanks}(1996)]{CroomShanks1996}
{Croom} S.~M., {Shanks} T., 1996, \mnras, 281, 893

\bibitem[{Croton}(2009)]{Croton2009}
{Croton} D.~J., 2009, \mnras, 394, 1109

\bibitem[{da {\^A}ngela} et~al.(2008){da {\^A}ngela}, {Shanks}, {Croom}
  et~al.]{daAngela2008}
{da {\^A}ngela} J., {Shanks} T., {Croom} S.~M., et~al., 2008, \mnras, 383, 565

\bibitem[{Degraf} et~al.(2010){Degraf}, {Di Matteo} \& {Springel}]{DeGraf2010}
{Degraf} C., {Di Matteo} T., {Springel} V., 2010, \mnras, 402, 1927

\bibitem[{Di Matteo} et~al.(2008){Di Matteo}, {Colberg}, {Springel},
  {Hernquist} \& {Sijacki}]{DiMatteo2008}
{Di Matteo} T., {Colberg} J., {Springel} V., {Hernquist} L., {Sijacki} D.,
  2008, \apj, 676, 33

\bibitem[{Di Matteo} et~al.(2005){Di Matteo}, {Springel} \&
  {Hernquist}]{2005Natur...433..604D}
{Di Matteo} T., {Springel} V., {Hernquist} L., 2005, Nature, 433, 604

\bibitem[{Djorgovski}(1991)]{Djorgovski1991}
{Djorgovski} S., 1991, in { The Space Distribution of Quasars\/}, edited by
  {D.~Crampton}, vol.~21 of { Astronomical Society of the Pacific Conference
  Series\/},  349--353

\bibitem[{Ferrarese} \& {Merritt}(2000)]{2000ApJ...539L...9F}
{Ferrarese} L., {Merritt} D., 2000, \apjl, 539, L9

\bibitem[{Gebhardt} et~al.(2000){Gebhardt}, {Bender}, {Bower}
  et~al.]{2000ApJ...539L..13G}
{Gebhardt} K., {Bender} R., {Bower} G., et~al., 2000, \apjl, 539, L13

\bibitem[{Gerke} et~al.(2007){Gerke}, {Newman}, {Lotz} et~al.]{Gerke2007}
{Gerke} B.~F., {Newman} J.~A., {Lotz} J., et~al., 2007, \apjl, 660, L23

\bibitem[{Graham} \& {Driver}(2007)]{2007ApJ...655...77G}
{Graham} A.~W., {Driver} S.~P., 2007, \apj, 655, 77

\bibitem[{Haiman} \& {Hui}(2001)]{HaimanHui2001}
{Haiman} Z., {Hui} L., 2001, \apj, 547, 27

\bibitem[{Hennawi} et~al.(2006){Hennawi}, {Strauss}, {Oguri}
  et~al.]{Hennawi2006}
{Hennawi} J.~F., {Strauss} M.~A., {Oguri} M., et~al., 2006, AJ, 131, 1

\bibitem[{Hewett} et~al.(1998){Hewett}, {Foltz}, {Harding} \&
  {Lewis}]{Hewett1998}
{Hewett} P.~C., {Foltz} C.~B., {Harding} M.~E., {Lewis} G.~F., 1998, AJ, 115,
  383

\bibitem[{Hopkins} et~al.(2005{\natexlab{a}}){Hopkins}, {Hernquist}, {Cox}
  et~al.]{2005ApJ...630..705H}
{Hopkins} P.~F., {Hernquist} L., {Cox} T.~J., et~al., 2005{\natexlab{a}}, \apj,
  630, 705

\bibitem[{Hopkins} et~al.(2005{\natexlab{b}}){Hopkins}, {Hernquist}, {Cox}, {Di
  Matteo}, {Robertson} \& {Springel}]{2005ApJ...630..716H}
{Hopkins} P.~F., {Hernquist} L., {Cox} T.~J., {Di Matteo} T., {Robertson} B.,
  {Springel} V., 2005{\natexlab{b}}, \apj, 630, 716

\bibitem[{Hopkins} et~al.(2005{\natexlab{c}}){Hopkins}, {Hernquist}, {Cox}, {Di
  Matteo}, {Robertson} \& {Springel}]{2005ApJ...632...81H}
{Hopkins} P.~F., {Hernquist} L., {Cox} T.~J., {Di Matteo} T., {Robertson} B.,
  {Springel} V., 2005{\natexlab{c}}, \apj, 632, 81

\bibitem[{Hopkins} et~al.(2006){Hopkins}, {Hernquist}, {Cox}, {Di Matteo},
  {Robertson} \& {Springel}]{2006ApJS..163....1H}
{Hopkins} P.~F., {Hernquist} L., {Cox} T.~J., {Di Matteo} T., {Robertson} B.,
  {Springel} V., 2006, \apjs, 163, 1

\bibitem[{Hopkins} et~al.(2005{\natexlab{d}}){Hopkins}, {Hernquist}, {Martini}
  et~al.]{2005ApJ...625L..71H}
{Hopkins} P.~F., {Hernquist} L., {Martini} P., et~al., 2005{\natexlab{d}},
  \apjl, 625, L71

\bibitem[{Hoyle} \& {Lyttleton}(1939)]{1939PCPS...35..405H}
{Hoyle} F., {Lyttleton} R.~A., 1939, in { Proceedings of the Cambridge
  Philosophical Society\/}, vol.~35 of { Proceedings of the Cambridge
  Philosophical Society\/},  405

\bibitem[{Kerscher} et~al.(2000){Kerscher}, {Szapudi} \&
  {Szalay}]{Kerscher2000}
{Kerscher} M., {Szapudi} I., {Szalay} A.~S., 2000, \apjl, 535, L13

\bibitem[{Kochanek} et~al.(1999){Kochanek}, {Falco} \&
  {Mu{\~n}oz}]{Kochanek1999}
{Kochanek} C.~S., {Falco} E.~E., {Mu{\~n}oz} J.~A., 1999, \apj, 510, 590

\bibitem[{Komossa} et~al.(2003){Komossa}, {Burwitz}, {Hasinger}, {Predehl},
  {Kaastra} \& {Ikebe}]{Komossa2003}
{Komossa} S., {Burwitz} V., {Hasinger} G., {Predehl} P., {Kaastra} J.~S.,
  {Ikebe} Y., 2003, \apjl, 582, L15

\bibitem[{Kormendy} \& {Richstone}(1995)]{1995ARA&A..33..581K}
{Kormendy} J., {Richstone} D., 1995, ARA\&A, 33, 581

\bibitem[{Kundic}(1997)]{Kundic1997}
{Kundic} T., 1997, \apj, 482, 631

\bibitem[{La Franca} et~al.(1998){La Franca}, {Andreani} \&
  {Cristiani}]{LaFranca1998}
{La Franca} F., {Andreani} P., {Cristiani} S., 1998, \apj, 497, 529

\bibitem[{Landy} \& {Szalay}(1993)]{LandySzalay1993}
{Landy} S.~D., {Szalay} A.~S., 1993, \apj, 412, 64

\bibitem[{Lewis} et~al.(2002){Lewis}, {Cannon}, {Taylor}
  et~al.]{2002MNRAS.333..279L}
{Lewis} I.~J., {Cannon} R.~D., {Taylor} K., et~al., 2002, \mnras, 333, 279

\bibitem[{Lidz} et~al.(2006){Lidz}, {Hopkins}, {Cox}, {Hernquist} \&
  {Robertson}]{Lidz2006}
{Lidz} A., {Hopkins} P.~F., {Cox} T.~J., {Hernquist} L., {Robertson} B., 2006,
  \apj, 641, 41

\bibitem[{Magliocchetti} \& {Porciani}(2003)]{Magliocchetti2003}
{Magliocchetti} M., {Porciani} C., 2003, \mnras, 346, 186

\bibitem[{Magorrian} et~al.(1998){Magorrian}, {Tremaine}, {Richstone}
  et~al.]{1998AJ....115.2285M}
{Magorrian} J., {Tremaine} S., {Richstone} D., et~al., 1998, AJ, 115, 2285

\bibitem[{Martini} \& {Weinberg}(2001)]{MartiniWeinberg2001}
{Martini} P., {Weinberg} D.~H., 2001, \apj, 547, 12

\bibitem[{Mo} \& {Fang}(1993)]{MoFang1993}
{Mo} H.~J., {Fang} L.~Z., 1993, \apj, 410, 493

\bibitem[{Mo} \& {White}(2002)]{MoWhite2002}
{Mo} H.~J., {White} S.~D.~M., 2002, \mnras, 336, 112

\bibitem[{Mortlock} et~al.(1999){Mortlock}, {Webster} \&
  {Francis}]{Mortlock1999}
{Mortlock} D.~J., {Webster} R.~L., {Francis} P.~J., 1999, \mnras, 309, 836

\bibitem[{Myers} et~al.(2007{\natexlab{a}}){Myers}, {Brunner}, {Nichol},
  {Richards}, {Schneider} \& {Bahcall}]{Myers2007}
{Myers} A.~D., {Brunner} R.~J., {Nichol} R.~C., {Richards} G.~T., {Schneider}
  D.~P., {Bahcall} N.~A., 2007{\natexlab{a}}, \apj, 658, 85

\bibitem[{Myers} et~al.(2007{\natexlab{b}}){Myers}, {Brunner}, {Richards},
  {Nichol}, {Schneider} \& {Bahcall}]{Myers2007II}
{Myers} A.~D., {Brunner} R.~J., {Richards} G.~T., {Nichol} R.~C., {Schneider}
  D.~P., {Bahcall} N.~A., 2007{\natexlab{b}}, \apj, 658, 99

\bibitem[{Myers} et~al.(2008){Myers}, {Richards}, {Brunner} et~al.]{Myers2008}
{Myers} A.~D., {Richards} G.~T., {Brunner} R.~J., et~al., 2008, \apj, 678, 635

\bibitem[{Navarro} et~al.(1996){Navarro}, {Frenk} \& {White}]{NFW1996}
{Navarro} J.~F., {Frenk} C.~S., {White} S.~D.~M., 1996, \apj, 462, 563

\bibitem[{Padmanabhan} et~al.(2009){Padmanabhan}, {White}, {Norberg} \&
  {Porciani}]{Padmanabhan2009}
{Padmanabhan} N., {White} M., {Norberg} P., {Porciani} C., 2009, \mnras, 397,
  1862

\bibitem[{Peacock}(1999)]{Peacock1999}
{Peacock} J.~A., 1999, {Cosmological Physics}, Cosmological Physics, by John
  A.~Peacock, pp.~704.~ISBN 052141072X.~Cambridge, UK: Cambridge University
  Press, January 1999.

\bibitem[{Porciani} et~al.(2004){Porciani}, {Magliocchetti} \&
  {Norberg}]{Porciani2004}
{Porciani} C., {Magliocchetti} M., {Norberg} P., 2004, \mnras, 355, 1010

\bibitem[{Ross} et~al.(2009){Ross}, {Shen}, {Strauss} et~al.]{Ross2009}
{Ross} N.~P., {Shen} Y., {Strauss} M.~A., et~al., 2009, \apj, 697, 1634

\bibitem[{Shakura} \& {Sunyaev}(1973)]{1973A&A....24..337S}
{Shakura} N.~I., {Sunyaev} R.~A., 1973, A\&A, 24, 337

\bibitem[{Shen} et~al.(2009{\natexlab{a}}){Shen}, {Hennawi}, {Shankar}
  et~al.]{Shen2009b}
{Shen} Y., {Hennawi} J.~F., {Shankar} F., et~al., 2009{\natexlab{a}}, ArXiv
  e-prints

\bibitem[{Shen} et~al.(2007){Shen}, {Strauss}, {Oguri} et~al.]{Shen2007}
{Shen} Y., {Strauss} M.~A., {Oguri} M., et~al., 2007, AJ, 133, 2222

\bibitem[{Shen} et~al.(2009{\natexlab{b}}){Shen}, {Strauss}, {Ross}
  et~al.]{Shen2009}
{Shen} Y., {Strauss} M.~A., {Ross} N.~P., et~al., 2009{\natexlab{b}}, \apj,
  697, 1656

\bibitem[{Springel} et~al.(2001){Springel}, {White}, {Tormen} \&
  {Kauffmann}]{Springel2001}
{Springel} V., {White} S.~D.~M., {Tormen} G., {Kauffmann} G., 2001, \mnras,
  328, 726

\bibitem[{Tremaine} et~al.(2002){Tremaine}, {Gebhardt}, {Bender}
  et~al.]{2002ApJ...574..740T}
{Tremaine} S., {Gebhardt} K., {Bender} R., et~al., 2002, \apj, 574, 740

\bibitem[{York} et~al.(2000){York}, {Adelman}, {Anderson}
  et~al.]{2000AJ....120.1579Y}
{York} D.~G., {Adelman} J., {Anderson} Jr. J.~E., et~al., 2000, AJ, 120, 1579

\bibitem[{Zehavi} et~al.(2004){Zehavi}, {Weinberg}, {Zheng} et~al.]{Zehavi2004}
{Zehavi} I., {Weinberg} D.~H., {Zheng} Z., et~al., 2004, \apj, 608, 16

\end{thebibliography}

\end{document}